\documentclass[aps,prl,10pt,twocolumn,superscriptaddress,nofootinbib,nobibnotes]{revtex4-1}

\usepackage[T1]{fontenc}
\usepackage[utf8]{inputenc}
\usepackage{amsmath}
\usepackage{amssymb}
\usepackage{graphicx}
\usepackage{siunitx}

\usepackage[x11names]{xcolor}

\usepackage{fancyhdr}
\newcommand{\stylecolor}{green!35!black}

\usepackage[explicit]{titlesec}

\usepackage[unicode=true,
            colorlinks=true,
            linkcolor=blue,
            citecolor=blue,
            urlcolor=blue]{hyperref} 
\usepackage{tikz}
\usepackage{isomath}

\setcitestyle{super}

\renewcommand{\d}[1]{\ensuremath{\mathop{{\rm d}{#1}}}}
\newcommand{\Vbias}{\ensuremath{V_{\rm b}}}

\begin{document}

\title{\Large \color{\stylecolor} \textsf{Thermodynamic reciprocity in scanning photocurrent maps}}


\author{Mark B. Lundeberg}
\email{mark.lundeberg@gmail.com}
\affiliation{ICFO --- Institut de Ciències Fotòniques, The Barcelona Institute of Science and Technology, 08860 Castelldefels (Barcelona), Spain}
\author{Frank H. L. Koppens}
\affiliation{ICFO --- Institut de Ciències Fotòniques, The Barcelona Institute of Science and Technology, 08860 Castelldefels (Barcelona), Spain}
\affiliation{ICREA --- Institució Catalana de Recerça i Estudis Avancats, Barcelona, Spain.}



\begin{abstract}
Scanning photocurrent maps in inhomogeneous materials contain nontrivial patterns, which often can only be understood with a full model of device geometry and nonuniformities.
We remark on the consequences of Onsager reciprocity to the photocurrent in linear response, with immediate applications in photovoltaic and photothermoelectric effects.
In particular with photothermoelectric effects, we find that the ampere-per-watt responsivity is exactly governed by Peltier-induced temperature shifts in the same device when time-reversed and voltage-biased.
We show, with the example of graphene, that this principle aids in understanding and modelling of photocurrent maps.
\end{abstract}


\maketitle

\titleformat{\section}
  {\gdef\sectionlabel{}
   \large\bfseries\scshape}
  {\gdef\sectionlabel{\thesection }}{0pt}
  {\begin{tikzpicture}[remember picture,overlay]
	\draw (1, 0) node[right] {\color{\stylecolor} \textsf{#1}};
	\fill[color=\stylecolor] (0,-0.25) rectangle (0.7, 0.25);
	\draw (0.35, 0) node {\color{white} \textsf{\sectionlabel}};
       \end{tikzpicture}
  }
\titlespacing*{\section}{0pt}{10pt}{10pt}

\section{Introduction}

A photocurrent map is the result of scanning a focussed beam of light over an optoelectronic device, recording the output current for each focus position.
These maps can yield crucial information about the local physics of current generation.
Unfortunately, the local physics is not the only factor determining the photocurrent: the {\em measured} current, passing through an ammeter often located meters away, is influenced also by the greater device geometry, inhomogeneities, electrode configuration, and external circuitry.
The quantitative explanation of photocurrent thus typically requires a full-device simulation.
In principle, then, modelling a photocurrent map is a numerically intensive task, since the entire device must be repeatedly simulated for each light focus position.
Such complex simulations cost time, and can obscure the understanding of simple and essential effects, providing no intuitive guidelines towards optimization.

In this manuscript, we provide a general Onsager reciprocity approach to link photocurrent maps to the behaviour of the device when voltage biased and un-illuminated.
The maps are thus computable in a single simulation, and directly related to a transport problem which may have a more intuitive solution.
Our results are restricted to linear and near-equilibrium photoresponse, but are otherwise very general as they are not based on any particular microscopic transport model.
This generalizes similar reciprocity techniques such as minority-carrier reciprocity\cite{DelAlamo1984,Donolato1985,Misiakos1985,Markvart1996,Green1997,Rau1998} or the Shockley--Ramo theorem,\cite{Shockley1938,Ramo1939,Song2014} which previously were justified on the basis of particular transport models.
We furthermore apply this approach to find an application in photothermoelectric effects, and investigate it in detail by way of an example.

\section{Context}

Herein, we assume that the direct effect of light absorption is to inject some thermodynamic components, such as charge or energy, into a device.
Let the injected flow of component $i$ into location $\vec x$ be represented by a injection density $J_i(\vec x)$.
An external current $I$ is measured via an attached drain electrode.

The quantity of interest---photocurrent---is the external current induced by light.
Since we have assumed that the effect of the light is fully captured in $J_i(\vec x)$, then the only possible form for the photocurrent in linear response is a superposition, i.e., a weighted volume integral
\begin{align}
 I = \sum_i\iint \d{\vec x} \, r_i(\vec x) J_i(\vec x)
 ,
 \label{eq:integral-general}
\end{align}
which defines a coefficient $r_i(\vec x)$, representing the internal responsivity of the device to a quantity $i$ injected at location $\vec x$.

The utility of Eq.~\eqref{eq:integral-general} is illustrated with the following example.
Consider that the incident light is tightly focussed at location $\vec x_{\rm c}$ and simply heats the device at that point, so that the only relevant component is injected energy $J_E(\vec x) = \alpha P \delta(\vec x-\vec x_{\rm c})$, for absorption efficiency $\alpha$ and incident power $P$.
The photocurrent map (dependence of $I$ on scanned $\vec x_{\rm c}$) is then given by $I(\vec x_{\rm c}) = \alpha P r_E(\vec x_{\rm c})$.
Alternatively if the light has a large focus, the photocurrent map will appear as a smoothed-out version of $\alpha P r_E(\vec x)$.
Generally, $J_i(\vec x)$ is dependent on the local electrodynamics and material-specific physics of light absorption, which are beyond the scope of this work.
Our concern here is determination of $r_i(\vec x)$, which will depend nonlocally on geometry, external circuitry, device inhomogeneities, and so on.

\section{Onsager theory}

Having established the utility of $r_i(\vec x)$, we now show how it can be simply determined by modelling the device under a bias $V_{\rm b}$ applied at the drain electrode.
Consider in isolation a single infinitesimal element $\d{\vec x}$ of Eq.~\eqref{eq:integral-general}, for some particular $i, \vec x$.
As a trick, we re-express the injection process by a connection from $\vec x$ to a virtual external thermodynamic $i$-reservoir, labelled ``a''.
To drive the correct $i$-current in this connection, $I_{\rm a} = - J_i(\vec x) \d{\vec x}$, we impose an appropriate thermodynamic state in the reservoir, out of balance with the device at location $\vec x$.
A second thermodynamic reservoir represents the electrode, which carries charge current $I_{\rm b} = I$ to a charge reservoir labelled ``b''.
This trick reinterprets the optoelectronic device as a black-box multi-terminal carrier of thermodynamic flows, to which we can apply universal multi-terminal reciprocity rules.

To express the reservoir states in a convenient way, we use thermodynamic affinities $f_{\rm a}$ and $f_{\rm b}$.
Here an {\em affinity} is defined as the intensive variable that is entropically conjugate to an extensive quantity: energy affinity is $1/T$ and charge affinity is $-V/T$, where $T$ is temperature and $V$ is voltage.
When only these two external reservoirs A and B are disturbed from equilibrium, with small deviations $\delta f_{\rm a}$ and $\delta f_{\rm b}$, the currents are related by a matrix:
\begin{align}
\begin{bmatrix}
 I_{\rm a} \\ I_{\rm b}
\end{bmatrix}
=
\begin{bmatrix}
 L_{\rm aa} & L_{\rm ab} \\
 L_{\rm ba} & L_{\rm bb} \\
\end{bmatrix}
\begin{bmatrix}
 \delta f_{\rm a} \\ \delta f_{\rm b}
\end{bmatrix}
,
\label{eq:onsager}
\end{align}
for some coefficients $L_{ij}$ (the values of $L_{ij}$ are not important, only the relationships between them as we use below).

By Onsager's reciprocity principle,\cite{Onsager1931,Casimir1945} the $L$ matrix is transposed upon time reversal of the system: $L_{ij}(+\vec B) = L_{ji}(-\vec B)$
[we use the label ``$-\vec B$'' to refer to time-reversal since it often requires only reversing the magnetic field].
This implies the following connection:
\begin{align}
\begin{split}
 \left(\frac{I_{\rm b}}{I_{\rm a}}\right)_{\delta f_{\rm b} = 0, +\vec B}
 & = \left(\frac{L_{\rm ba}}{L_{\rm aa}}\right)_{+\vec B} \\
 & = \left(\frac{L_{\rm ab}}{L_{\rm aa}}\right)_{-\vec B} = -\left(\frac{\delta f_{\rm a}}{\delta f_{\rm b}}\right)_{I_{\rm a} = 0, -\vec B}
.
\end{split}
\label{eq:gen-AB}
\end{align}
The significance of this becomes clear after substituting the appropriate values.
Note that $\delta f_{\rm b} = -V_{\rm b}/T$.
Crucially in the RHS, in order to have zero current in connection ``a'', its ends must have equal affinities.
Thus, $\delta f_{\rm a} = \delta f_i(\vec x)$, where $\delta f_i(\vec x)$ is the affinity in the device at the point $\vec x$ of connection.
Equation~\eqref{eq:gen-AB} becomes:
\begin{align}
\left(\frac{I}{-J_i(\vec x) \d{\vec x}}\right)_{V_{\rm b} = 0,+\vec B}
& = -\left(\frac{\delta f_i(\vec x)}{-V_{\rm b}/T}\right)_{J_i=0,-\vec B},
\label{eq:gen}
\end{align}
where it can now be seen that the LHS is the photocurrent from $J_i(\vec x)$, while the RHS is the bias-response without illumination ($J_i=0$).
This result is seemingly independent of our virtual-connection trick, however it does require a well-defined local $\delta f_i(\vec x)$ to which an unambiguous thermodynamic connection could be made.

Returning to Eq.~\eqref{eq:integral-general}, we thus have
\begin{align}
 r_i(\vec x) =  -T\left( \frac{\delta f_i(\vec x)}{\Vbias} \right)_{{\rm bias},-\vec B}
 .
 \label{eq:integral-coefficient}
\end{align}
In other words the responsivity coefficient---and hence photocurrent map---is governed by the internal landscape of affinity shifts induced by bias voltages, in the time-reversed system.
(A similar analysis yields the photo{\em voltage} responsivity in terms of current-biased response.)
Due to the flexibility of what component $i$ may represent, this reciprocity principle can encapsulate a wide range of photocurrent phenomena.

\section{Applications}

Equations \eqref{eq:integral-general} and \eqref{eq:integral-coefficient} form a general framework, which we now apply to three different cases of interest.

\begin{figure}
\raisebox{-.5\height}{\includegraphics[page=2]{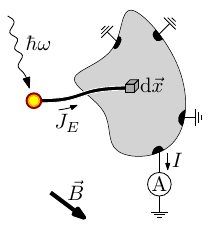}}%
~~~~{\Large$ \Leftrightarrow$}~~~~%
\raisebox{-.5\height}{\includegraphics[page=3]{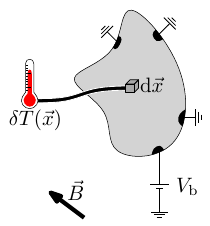}}
 \caption{
 \label{fig:topology}
Schematic describing reciprocity as it applies to photothermoelectric effects, Eqs.~\eqref{eq:principle-heat-usage} and \eqref{eq:principle-heat-reciprocity}.
Suppose the effect of light absorption can be represented as a local heat injector, resulting in a measured external current (left).
The response is reciprocal and equivalent to measuring the local temperature response to an applied voltage, without light (right).
}
\end{figure}

{\em Photothermoelectric}---%
Light absorption heats a material, and in metals and semimetals the resulting thermoelectric effect typically provides the dominant photoresponse.
Consider that $J_{E}(\vec x)$ is an energy injection (``heating'') distribution, with corresponding energy affinity shift $\delta f_{E}(\vec x) = -\delta T(\vec x)/T^2$.
Then,
\begin{equation}
I = \iint \d{\vec x} \, r_E(\vec x) J_{E}(\vec x),
\label{eq:principle-heat-usage}
\end{equation}
with responsivity coefficient via Eq.~\eqref{eq:integral-coefficient},
\begin{equation}
r_E(\vec x) = \frac{1}{T} \left( \frac{\delta T(\vec x)}{\Vbias} \right)_{{\rm bias},-\vec B}.
\label{eq:principle-heat-reciprocity}
\end{equation}
To be clear, here $\delta T(\vec x)$ is the local temperature change in linear response to bias; this includes linear Peltier effects but not Joule heating which is quadratic in bias.
Figure~\ref{fig:topology} illustrates this reciprocity principle.
Equations \eqref{eq:principle-heat-usage} and \eqref{eq:principle-heat-reciprocity} are a novel result and will be demonstrated in the following section.

{\em Photovoltaic}---%
In semiconductors and other multi-band conductors, interband photovoltaic effects dominate.
It is standard to separate the transport in the valence and conduction bands,\cite{Rau1998} having distinct voltages $V_{\rm V}$ and $V_{\rm C}$ (distinct quasi-Fermi levels) when out of equilibrium.
In this case the light induces electron-hole generation, i.e., injects a charge current $J_{\rm V}(\vec x) = +J_{\rm VC}(\vec x)$ into the valence band, and $J_{\rm C}(\vec x) = -J_{\rm VC}(\vec x)$ into the conduction band.
The charge affinities are $f_{\rm V}(\vec x) = -V_{\rm V}(\vec x)/T(\vec x)$ and $f_{\rm C}(\vec x) = -V_{\rm C}(\vec x)/T(\vec x)$, assuming the bands locally share temperature $T(\vec x)$.
Then, the photocurrent is given by
\begin{align}
 I & = \iint \d{\vec x} \, r_{\rm VC}(\vec x) J_{\rm VC}(\vec x),
\label{eq:PV-integral}
\end{align}
with responsivity coefficient via Eq.~\eqref{eq:integral-coefficient},
\begin{align}
 r_{\rm VC}(\vec x) & = \left(\frac{\delta V_{\rm V}(\vec x) - \delta V_{\rm C}(\vec x)}{\Vbias}\right)_{{\rm bias},-\vec B},
\label{eq:PV-reciprocity}
\end{align}
extending the Rau--Brendel reciprocity\cite{Rau1998} to nonzero magnetic field.
In lightly-doped semiconductors this $r_{\rm VC}(\vec x)$ can be related to the minority carrier concentration change under bias,\cite{DelAlamo1984,Donolato1985,Misiakos1985,Markvart1996,Green1997,Rau1998} however the more general Eq.~\eqref{eq:PV-reciprocity} does not make an arbitrary minority--majority carrier distinction.

More generally the light induces both effects in a semiconductor, $J_{\rm E} (\vec x)$ energy absorbed (thermoelectric effect) and $J_{\rm VC}(\vec x)$ interband current (photovoltaic effect), and the total response is then a sum of Eqs.~\eqref{eq:principle-heat-usage} and \eqref{eq:PV-integral}.
If $\hbar\omega$ is the photon energy and each absorbed photon transfers one electron from valence to conduction band, then $J_{\rm E} = (\hbar\omega/e) J_{\rm VC}$.
The number of extracted electrons per photon absorbed (IQE---internal quantum efficiency) is then
\begin{align}
\text{IQE}(\vec x) = (\hbar \omega/e) r_{\rm E}(\vec x) + r_{\rm VC}(\vec x).
\end{align}
In optimized solar cells the interband contribution dominates,\cite{Kettemann2002} however in degenerately doped semiconductors, low-dimensional semiconductors (2d or 1d), or disordered semiconductors the two contributions may be comparable.
Also, for increasing $\hbar\omega$ the first contribution may be enhanced to arbitrarily large values, whereas the second contribution is typically restricted to sub-unity values unless carrier multiplication occurs.

{\em Photogalvanic}---%
An alternative to the above is the Shockley--Ramo picture.\cite{Song2014}
In this case rather than considering the direct effects of energy injection and interband currents, one starts from an indirect effect of light which is an extraneous lateral charge current density $\vec j_{\rm ph}(\vec x)$ (this indirection can lead to technical complications as decribed in the following section).
This removes and redeposits charge in different locations, giving charge injection $J_Q(\vec x) = -\vec\nabla \cdot \vec j_{\rm ph}(\vec x)$.
For such charge injection we have
\begin{align}
 I & = \iint \d{\vec x} \, r_Q(\vec x) J_Q(\vec x),
 \label{eq:shockley-ramo}
\end{align}
with responsivity coefficient via Eq.~\eqref{eq:integral-coefficient},\footnote{
We have $\delta f_Q(\vec x) = -\delta V(\vec x)/T + (V/T^2)\delta T$, which yields Eq.~\eqref{eq:shockley-ramo-reciprocity} assuming zero equilibrium voltage $V$.
For $V\neq 0$ we obtain an apparently unphysical $V$-dependent term, however this is cancelled upon recognizing that $\vec j_{\rm ph}(\vec x)$ must carry an associated energy current $V\vec j_{\rm ph}(\vec x)$, which compensates total $I$ via Eq.~\eqref{eq:principle-heat-reciprocity}.
}
\begin{align}
 r_Q(\vec x) & = \left(\frac{\delta V(\vec x)}{\Vbias}\right)_{{\rm bias},-\vec B}.
 \label{eq:shockley-ramo-reciprocity}
\end{align}
This recovers the Song--Levitov result\cite{Song2014}---note that $I = \iint \d{\vec x} \, \vec j_{\rm ph}(\vec x) \cdot \vec\nabla r_Q(\vec x)$, via integration by parts.

\section{Example: photothermoelectric effect}

As far as we are aware, the photothermoelectric reciprocity in Eq.~\eqref{eq:principle-heat-reciprocity} is a new result, and we would like to demonstrate its application by example.
Examples of the charge-pumping reciprocity principles, Eqs.~\eqref{eq:PV-reciprocity} and \eqref{eq:shockley-ramo-reciprocity}, can be found in the literature.\cite{DelAlamo1984,Donolato1985,Misiakos1985,Markvart1996,Green1997,Rau1998,Song2014}

Graphene photocurrent is typically dominated by thermoelectric effects, and its photocurrent maps can be highly nontrivial due to inhomogeneities or magnetic field.\cite{Cao2016,Woessner2016}
As a model, we solve the local thermoelectric transport equations\cite{Landau1984} for charge current $\vec j_Q$ and energy current $\vec j_E$, in terms of voltage $V$ and temperature $T$:
\begin{align}
\begin{split}
\vec j_Q & = - \sigma \vec\nabla V - \sigma S \vec\nabla T, \\
\vec j_E & = - (\Pi + V)\sigma \vec\nabla V - (\kappa + (\Pi + V)\sigma S)\vec\nabla T,
\end{split}
\label{eq:te-currents}
\end{align}
for conductivity $\sigma$, Seebeck coefficient $S$, Peltier cofficient $\Pi$, and thermal conductivity $\kappa$.
In magnetic field, these coefficients are tensor-valued but for simplicity we will consider a zero-field example.
In steady state, charge and energy are conserved:
\begin{align}
\begin{split}
\vec\nabla \cdot \vec j_Q & = 0,\\
\vec\nabla \cdot \vec j_E & = J_E - g(T - T_0),
\label{eq:te-conservation}
\end{split}
\end{align}
for heat energy input $J_E$ (the light absorbed) and heat leakage coefficient $g$ (to substrate).
In linear response, we calculate the coefficients above for equilibrium voltage and temperature.
We then numerically solve the linearized equations, with appropriate boundary conditions, via a finite volume method solved by sparse matrix techniques.
The results have been confirmed via both the FiPy package\cite{FiPy2009} and a custom code.

\begin{figure}
\includegraphics{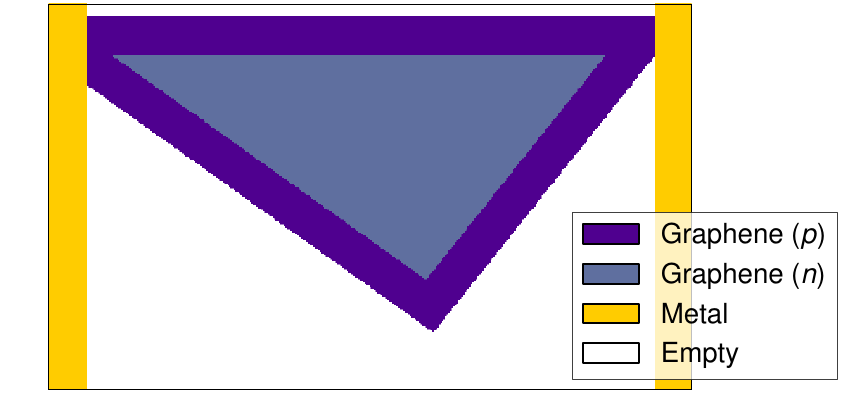}
 \caption{
 \label{fig:setup}
Example of 2D thermoelectric system involving a triangular, inhomogeneous graphene sheet, inspired by experiment.\cite{Woessner2016}
Plotted is a map of material types: a $p$-type perimeter surrounds a less conductive $n$-type core; metal electrodes make contact at two corners.
Current is measured at the right electrode while the left is grounded.
The system is discretized and solved on a $401\times 241$ square mesh.
}
\end{figure}

\begin{figure}
\includegraphics{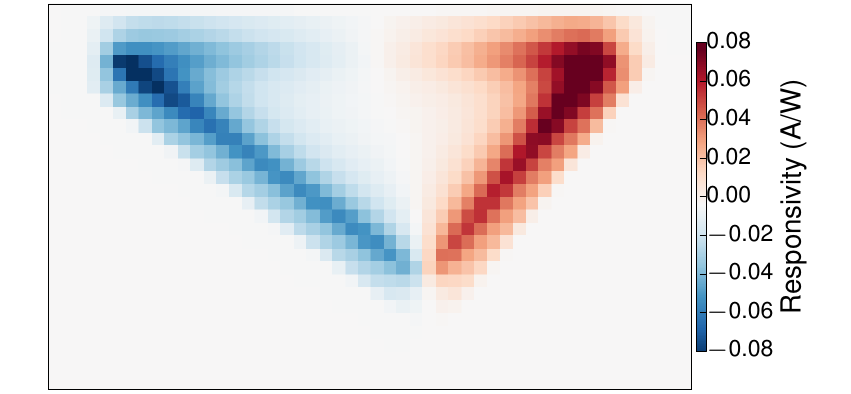}
 \caption{
 \label{fig:manualscan}
 Photocurrent map calculated directly, by injecting heat $J_E$ at various places and measuring current out the right electrode.
}
\end{figure}

\begin{figure}
\includegraphics{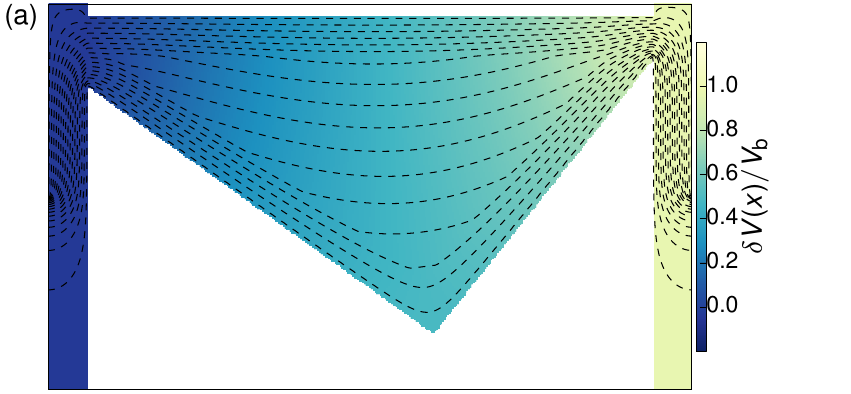}
\includegraphics{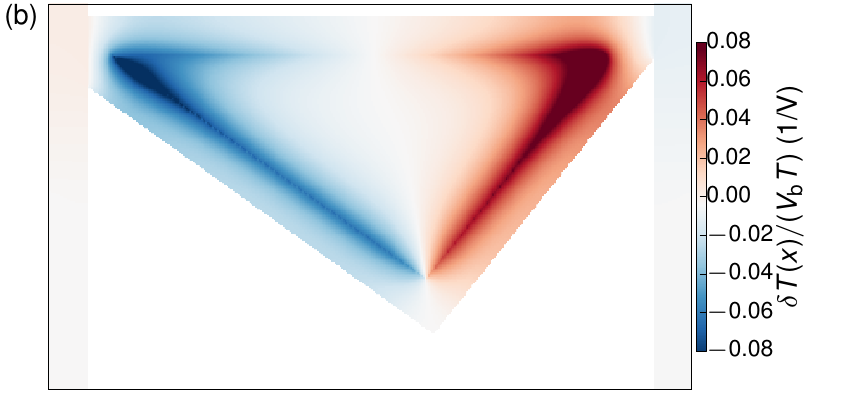}
 \caption{
 \label{fig:biased}
Solution with a voltage bias applied to the right electrode.
(a) A color plot showing the local voltage, scaled to the bias.
Dashed curves are the streamlines in $\vec j_Q$, spaced by equal flux.
(b) A color plot showing the temperature shifts, scaled to the bias and equilibrium temperature.
}
\end{figure}

Figures \ref{fig:setup}--\ref{fig:biased} examine a highly inhomogeneous case, representing a nonrectangular graphene device with carrier density variations.\cite{Woessner2016}
A triangular graphene sheet is contacted by electrodes at two corners: the left electrode is grounded, and the right electrode is grounded through an ammeter, measuring current $I$ and with possible bias voltage $V_{\rm b}$.
From the perimeter of the triangle to its core, the polarity of carrier concentration flips from n-type to p-type.
For brevity we omit the coefficients' values but note the salient details: that the core has lower values of $\sigma$, $\kappa$, and $g$, and that $\Pi$ and $S$ change sign from the perimeter to core (and indeed $\Pi = T S$ as we have taken zero magnetic field).

Figure \ref{fig:manualscan} shows the photocurrent map computed directly: for each pixel in the image, the device was simulated for a sharp $J_E(\vec x)$ distribution confined within the pixel, and $I$ was measured then normalized by the absorbed power $P = \iint \d{\vec x} \,J_E$.
The reciprocal setup is shown in Fig.~\ref{fig:biased}: we plot the internal landscape of $\delta V(\vec x)$ and $\delta T(\vec x)$ in the device, when driven by voltage bias.
The $\delta T$ map in Fig.~\ref{fig:biased} matches the photocurrent map in Fig.~\ref{fig:manualscan}---a consequence of reciprocity Eq.~\eqref{eq:principle-heat-reciprocity}.

We emphasize the computational speedup, that Fig.~\ref{fig:biased} required only one device simulation, whereas Fig.~\ref{fig:manualscan} required many repeated device simulations, one per pixel.
On a modern personal computer, each solution of this device (by sparse matrix techniques) required approximately one second, hence the time to obtain Fig.~\ref{fig:manualscan} was significant.
Moreover, the responsivity map in Fig.~\ref{fig:biased} is obtained with full mesh resolution, while in Fig.~\ref{fig:manualscan} the number of pixels was limited out of expediency.

It is worth pointing out the intuitive nature of the fields $\delta V(\vec x)$ and $\delta T(\vec x)$ in Fig.~\ref{fig:biased} (which, by reciprocity, exactly translate to intuitions about the photocurrent map).
First, note that the form of $\vec j_Q$, plotted as streamlines Fig.~\ref{fig:biased}(a), follows essentially the expected Ohmic flow of least resistance, tending to avoid the low-conductivity core region.
If $\vec j_Q$ is known then the temperature can be found from $\vec\nabla\cdot(\kappa\vec\nabla \delta T) - g\delta T = (\vec\nabla \Pi)\cdot \vec j_Q$, meaning that $\delta T(\vec x)$ is simply a diffused form of the Peltier heat.
The variations of Peltier heat along the junction, given by $(\Pi_1 - \Pi_2)\hat n \cdot \vec j_Q$ where $(\Pi_1-\Pi_2)$ is the step in Peltier coefficient, are determined by the variations in the current density $\vec j_Q$ relative to the junction normal vector $\hat n$.
After taking all of these factors together, with reciprocity, the responsivity of a straight junction far away from other inhomogeneities is
\begin{equation}
r_{\rm junc} \approx \frac{(\Pi_1 - \Pi_2)/T}{\sqrt{\kappa_1 g_1} + \sqrt{\kappa_2 g_2}} \bigg( \frac{\hat n \cdot \vec j_Q}{\Vbias} \bigg)_{{\rm bias}} ,
\label{eq:responsivity-junc}
\end{equation}
decaying away from the junction with the characteristic thermal length $l_{\rm th} = \sqrt{\kappa/g}$.
The variation in responsivity along the junction, as visible in Fig.~\ref{fig:biased}(b), has primarily to do with variations in $\hat n \cdot \vec j_Q$ as determined by geometry.
As a visual qualitative heuristic, $\hat n \cdot \vec j_Q$ is controlled by the density and direction with which current streamlines cross the the junction [Fig.~\ref{fig:biased}(a)], and accordingly it can be seen that regions of highest responsivity are those with highest density of streamlines crossing the junction.

We can compare to the intuitive picture of Song and Levitov\cite{Song2014}, where instead one considers how the heating $J_E$ causes a temperature rise over a localized region of size $l_{\rm th}$.
The resulting ``extraneous Seebeck current'' $\vec j_{\rm ph} = -\sigma S \vec\nabla T$ is on average directed along $\hat n$, and acts via Eq.~\eqref{eq:shockley-ramo} in combination with the bias-voltage profile (Fig.~\ref{fig:biased}) to give the photocurrent $I$.
This picture appears quite distinct but agrees on the approximate final photocurrent, Eq.~\eqref{eq:responsivity-junc}, since $\Pi = T S$ and the electrical transport is approximately Ohmic, $\vec\nabla V \approx \sigma^{-1} \vec j_Q$.

The thermoelectric equations~\eqref{eq:te-currents} may be strongly coupled in highly optimized devices, breaking these intuitions.
The bias-induced $\delta V$ and $\vec j_Q$ are no longer simple Ohmic conductivity solutions, becoming distorted by Peltier-Seebeck backaction.
Both Eqs.~\eqref{eq:principle-heat-usage} and \eqref{eq:shockley-ramo} then lose some intuitive value, though both remain technically correct.
In terms of computational utility, however, the two approaches differ: Eq.~\eqref{eq:principle-heat-usage} still provides a single-shot photocurrent map, whereas Eq.~\eqref{eq:shockley-ramo} does not.
This is because $J_E$ may induce temperature shifts at distant locations due to Seebeck-Peltier backaction, and if such backaction is significant then $\vec j_{\rm ph}$ can only be accurately determined by fully simulating the device.

\section{Conclusion}

Equation~\eqref{eq:integral-coefficient} has been shown to give remarkably direct proofs of Eqs.~\eqref{eq:PV-reciprocity} and \eqref{eq:shockley-ramo-reciprocity}, which were previously justified by careful examination of the Green functions of diffusive transport differential equations.
The thermoelectric reciprocity in Eq.~\eqref{eq:principle-heat-reciprocity} could also be justified by the microscopic Onsager relations\cite{Landau1984} between $\Pi$ and $S$ together with the Green functions of Eqs.~\eqref{eq:te-currents} and \eqref{eq:te-conservation}, however this is not necessary.
Nonlocal Onsager reciprocity---the statement that microscopic time reversal symmetry pervades thermodynamic transport at all scales---allows to bypass this process.

We anticipate that other photocurrent mechanisms are accessible with this method, dealing with other components besides charge and energy.
For example, in hydrodynamic, spintronic or valleytronic systems, it can be necessary to also consider the absorbed angular momentum from light.\cite{McIver2011}
In photoelectrochemical systems, the light induces a local chemical imbalance.\cite{Rau1998}
The incident light mode itself can even be considered as a thermodynamic component of the device,\cite{Markvart2008} giving electroluminescent reciprocity.\cite{Rau2007}
In each case we expect that Eq.~\eqref{eq:integral-coefficient} provides a direct route to the relevant photocurrent reciprocity principle, even in magnetic fields: every optoelectronic device near equilibrium is reciprocally related to its time-reversed counterpart under bias.

\begin{acknowledgments}
We thank K.-J. Tielrooij and S. Castilla for careful reading of the manuscript. F.H.L.K. acknowledges  support from the Government of Spain (FIS2016-81044; Severo Ochoa CEX2019-000910-S), Fundació Cellex, Fundació Mir-Puig, and Generalitat de Catalunya (CERCA, AGAUR, SGR 1656). Furthermore, the research leading to these results has received funding from the European Union’s Horizon 2020  under grant agreement  no. 881603 (Graphene flagship Core3). This work was supported by the ERC TOPONANOP under grant agreement  726001.
\end{acknowledgments}


\begin{thebibliography}{19}%
\makeatletter
\providecommand \@ifxundefined [1]{%
 \@ifx{#1\undefined}
}%
\providecommand \@ifnum [1]{%
 \ifnum #1\expandafter \@firstoftwo
 \else \expandafter \@secondoftwo
 \fi
}%
\providecommand \@ifx [1]{%
 \ifx #1\expandafter \@firstoftwo
 \else \expandafter \@secondoftwo
 \fi
}%
\providecommand \natexlab [1]{#1}%
\providecommand \enquote  [1]{``#1''}%
\providecommand \bibnamefont  [1]{#1}%
\providecommand \bibfnamefont [1]{#1}%
\providecommand \citenamefont [1]{#1}%
\providecommand \href@noop [0]{\@secondoftwo}%
\providecommand \href [0]{\begingroup \@sanitize@url \@href}%
\providecommand \@href[1]{\@@startlink{#1}\@@href}%
\providecommand \@@href[1]{\endgroup#1\@@endlink}%
\providecommand \@sanitize@url [0]{\catcode `\\12\catcode `\$12\catcode
  `\&12\catcode `\#12\catcode `\^12\catcode `\_12\catcode `\%12\relax}%
\providecommand \@@startlink[1]{}%
\providecommand \@@endlink[0]{}%
\providecommand \url  [0]{\begingroup\@sanitize@url \@url }%
\providecommand \@url [1]{\endgroup\@href {#1}{\urlprefix }}%
\providecommand \urlprefix  [0]{URL }%
\providecommand \Eprint [0]{\href }%
\providecommand \doibase [0]{http://dx.doi.org/}%
\providecommand \selectlanguage [0]{\@gobble}%
\providecommand \bibinfo  [0]{\@secondoftwo}%
\providecommand \bibfield  [0]{\@secondoftwo}%
\providecommand \translation [1]{[#1]}%
\providecommand \BibitemOpen [0]{}%
\providecommand \bibitemStop [0]{}%
\providecommand \bibitemNoStop [0]{.\EOS\space}%
\providecommand \EOS [0]{\spacefactor3000\relax}%
\providecommand \BibitemShut  [1]{\csname bibitem#1\endcsname}%
\let\auto@bib@innerbib\@empty
\bibitem [{\citenamefont {Del~Alamo}\ and\ \citenamefont
  {Swanson}(1984)}]{DelAlamo1984}%
  \BibitemOpen
  \bibfield  {author} {\bibinfo {author} {\bibfnamefont {J.~A.}\ \bibnamefont
  {Del~Alamo}}\ and\ \bibinfo {author} {\bibfnamefont {R.~M.}\ \bibnamefont
  {Swanson}},\ }\href {\doibase 10.1109/T-ED.1984.21805} {\bibfield  {journal}
  {\bibinfo  {journal} {IEEE Trans. Electron Dev.}\ }\textbf {\bibinfo {volume}
  {31}},\ \bibinfo {pages} {1878} (\bibinfo {year} {1984})}\BibitemShut
  {NoStop}%
\bibitem [{\citenamefont {{Donolato}}(1985)}]{Donolato1985}%
  \BibitemOpen
  \bibfield  {author} {\bibinfo {author} {\bibfnamefont {C.}~\bibnamefont
  {{Donolato}}},\ }\href {\doibase 10.1063/1.95654} {\bibfield  {journal}
  {\bibinfo  {journal} {Appl. Phys. Lett.}\ }\textbf {\bibinfo {volume} {46}},\
  \bibinfo {pages} {270} (\bibinfo {year} {1985})}\BibitemShut {NoStop}%
\bibitem [{\citenamefont {{Misiakos}}\ and\ \citenamefont
  {{Lindholm}}(1985)}]{Misiakos1985}%
  \BibitemOpen
  \bibfield  {author} {\bibinfo {author} {\bibfnamefont {K.}~\bibnamefont
  {{Misiakos}}}\ and\ \bibinfo {author} {\bibfnamefont {F.~A.}\ \bibnamefont
  {{Lindholm}}},\ }\href {\doibase 10.1063/1.336226} {\bibfield  {journal}
  {\bibinfo  {journal} {J. Appl. Phys.}\ }\textbf {\bibinfo {volume} {58}},\
  \bibinfo {pages} {4743} (\bibinfo {year} {1985})}\BibitemShut {NoStop}%
\bibitem [{\citenamefont {{Markvart}}(1996)}]{Markvart1996}%
  \BibitemOpen
  \bibfield  {author} {\bibinfo {author} {\bibfnamefont {T.}~\bibnamefont
  {{Markvart}}},\ }\href {\doibase 10.1109/16.502143} {\bibfield  {journal}
  {\bibinfo  {journal} {IEEE Trans. Electron Dev.}\ }\textbf {\bibinfo {volume}
  {43}},\ \bibinfo {pages} {1034} (\bibinfo {year} {1996})}\BibitemShut
  {NoStop}%
\bibitem [{\citenamefont {{Green}}(1997)}]{Green1997}%
  \BibitemOpen
  \bibfield  {author} {\bibinfo {author} {\bibfnamefont {M.~A.}\ \bibnamefont
  {{Green}}},\ }\href {\doibase 10.1063/1.364108} {\bibfield  {journal}
  {\bibinfo  {journal} {J. Appl. Phys.}\ }\textbf {\bibinfo {volume} {81}},\
  \bibinfo {pages} {268} (\bibinfo {year} {1997})}\BibitemShut {NoStop}%
\bibitem [{\citenamefont {Rau}\ and\ \citenamefont {Brendel}(1998)}]{Rau1998}%
  \BibitemOpen
  \bibfield  {author} {\bibinfo {author} {\bibfnamefont {U.}~\bibnamefont
  {Rau}}\ and\ \bibinfo {author} {\bibfnamefont {R.}~\bibnamefont {Brendel}},\
  }\href {\doibase 10.1063/1.368968} {\bibfield  {journal} {\bibinfo  {journal}
  {J. Appl. Phys.}\ }\textbf {\bibinfo {volume} {84}},\ \bibinfo {pages} {6412}
  (\bibinfo {year} {1998})}\BibitemShut {NoStop}%
\bibitem [{\citenamefont {Shockley}(1938)}]{Shockley1938}%
  \BibitemOpen
  \bibfield  {author} {\bibinfo {author} {\bibfnamefont {W.}~\bibnamefont
  {Shockley}},\ }\href {\doibase 10.1063/1.1710367} {\bibfield  {journal}
  {\bibinfo  {journal} {J. Appl. Phys.}\ }\textbf {\bibinfo {volume} {9}},\
  \bibinfo {pages} {635} (\bibinfo {year} {1938})}\BibitemShut {NoStop}%
\bibitem [{\citenamefont {Ramo}(1939)}]{Ramo1939}%
  \BibitemOpen
  \bibfield  {author} {\bibinfo {author} {\bibfnamefont {S.}~\bibnamefont
  {Ramo}},\ }\href {\doibase 10.1109/jrproc.1939.228757} {\bibfield  {journal}
  {\bibinfo  {journal} {Proc. IRE}\ }\textbf {\bibinfo {volume} {27}},\
  \bibinfo {pages} {584–585} (\bibinfo {year} {1939})}\BibitemShut {NoStop}%
\bibitem [{\citenamefont {Song}\ and\ \citenamefont
  {Levitov}(2014)}]{Song2014}%
  \BibitemOpen
  \bibfield  {author} {\bibinfo {author} {\bibfnamefont {J.~C.~W.}\
  \bibnamefont {Song}}\ and\ \bibinfo {author} {\bibfnamefont {L.~S.}\
  \bibnamefont {Levitov}},\ }\href {\doibase 10.1103/PhysRevB.90.075415}
  {\bibfield  {journal} {\bibinfo  {journal} {Phys. Rev. B}\ }\textbf {\bibinfo
  {volume} {90}},\ \bibinfo {pages} {075415} (\bibinfo {year}
  {2014})}\BibitemShut {NoStop}%
\bibitem [{\citenamefont {Onsager}(1931)}]{Onsager1931}%
  \BibitemOpen
  \bibfield  {author} {\bibinfo {author} {\bibfnamefont {L.}~\bibnamefont
  {Onsager}},\ }\href {\doibase 10.1103/physrev.37.405} {\bibfield  {journal}
  {\bibinfo  {journal} {Phys. Rev.}\ }\textbf {\bibinfo {volume} {37}},\
  \bibinfo {pages} {405–426} (\bibinfo {year} {1931})}\BibitemShut {NoStop}%
\bibitem [{\citenamefont {Casimir}(1945)}]{Casimir1945}%
  \BibitemOpen
  \bibfield  {author} {\bibinfo {author} {\bibfnamefont {H.~B.~G.}\
  \bibnamefont {Casimir}},\ }\href {\doibase 10.1103/RevModPhys.17.343}
  {\bibfield  {journal} {\bibinfo  {journal} {Rev. Mod. Phys.}\ }\textbf
  {\bibinfo {volume} {17}},\ \bibinfo {pages} {343} (\bibinfo {year}
  {1945})}\BibitemShut {NoStop}%
\bibitem [{\citenamefont {Kettemann}\ and\ \citenamefont
  {Guillemoles}(2002)}]{Kettemann2002}%
  \BibitemOpen
  \bibfield  {author} {\bibinfo {author} {\bibfnamefont {S.}~\bibnamefont
  {Kettemann}}\ and\ \bibinfo {author} {\bibfnamefont {J.-F.}\ \bibnamefont
  {Guillemoles}},\ }\href {\doibase 10.1016/s1386-9477(02)00365-x} {\bibfield
  {journal} {\bibinfo  {journal} {Physica E}\ }\textbf {\bibinfo {volume}
  {14}},\ \bibinfo {pages} {101} (\bibinfo {year} {2002})}\BibitemShut
  {NoStop}%
\bibitem [{\citenamefont {{Cao}}\ \emph {et~al.}(2016)\citenamefont {{Cao}},
  \citenamefont {{Aivazian}}, \citenamefont {{Fei}}, \citenamefont {{Ross}},
  \citenamefont {{Cobden}},\ and\ \citenamefont {{Xu}}}]{Cao2016}%
  \BibitemOpen
  \bibfield  {author} {\bibinfo {author} {\bibfnamefont {H.}~\bibnamefont
  {{Cao}}}, \bibinfo {author} {\bibfnamefont {G.}~\bibnamefont {{Aivazian}}},
  \bibinfo {author} {\bibfnamefont {Z.}~\bibnamefont {{Fei}}}, \bibinfo
  {author} {\bibfnamefont {J.}~\bibnamefont {{Ross}}}, \bibinfo {author}
  {\bibfnamefont {D.~H.}\ \bibnamefont {{Cobden}}}, \ and\ \bibinfo {author}
  {\bibfnamefont {X.}~\bibnamefont {{Xu}}},\ }\href {\doibase
  10.1038/nphys3549} {\bibfield  {journal} {\bibinfo  {journal} {Nat. Phys.}\
  }\textbf {\bibinfo {volume} {12}},\ \bibinfo {pages} {236} (\bibinfo {year}
  {2016})}\BibitemShut {NoStop}%
\bibitem [{\citenamefont {Woessner}\ \emph {et~al.}(2016)\citenamefont
  {Woessner}, \citenamefont {Alonso-González}, \citenamefont {Lundeberg},
  \citenamefont {Gao}, \citenamefont {Barrios-Vargas}, \citenamefont
  {Navickaite}, \citenamefont {Ma}, \citenamefont {Janner}, \citenamefont
  {Watanabe}, \citenamefont {Cummings},\ and\ \citenamefont
  {et~al.}}]{Woessner2016}%
  \BibitemOpen
  \bibfield  {author} {\bibinfo {author} {\bibfnamefont {A.}~\bibnamefont
  {Woessner}}, \bibinfo {author} {\bibfnamefont {P.}~\bibnamefont
  {Alonso-González}}, \bibinfo {author} {\bibfnamefont {M.~B.}\ \bibnamefont
  {Lundeberg}}, \bibinfo {author} {\bibfnamefont {Y.}~\bibnamefont {Gao}},
  \bibinfo {author} {\bibfnamefont {J.~E.}\ \bibnamefont {Barrios-Vargas}},
  \bibinfo {author} {\bibfnamefont {G.}~\bibnamefont {Navickaite}}, \bibinfo
  {author} {\bibfnamefont {Q.}~\bibnamefont {Ma}}, \bibinfo {author}
  {\bibfnamefont {D.}~\bibnamefont {Janner}}, \bibinfo {author} {\bibfnamefont
  {K.}~\bibnamefont {Watanabe}}, \bibinfo {author} {\bibfnamefont {A.~W.}\
  \bibnamefont {Cummings}}, \ and\ \bibinfo {author} {\bibnamefont {et~al.}},\
  }\href {\doibase 10.1038/ncomms10783} {\bibfield  {journal} {\bibinfo
  {journal} {Nat Comms}\ }\textbf {\bibinfo {volume} {7}},\ \bibinfo {pages}
  {10783} (\bibinfo {year} {2016})}\BibitemShut {NoStop}%
\bibitem [{\citenamefont {Landau}\ \emph {et~al.}(1984)\citenamefont {Landau},
  \citenamefont {Bell}, \citenamefont {Kearsley}, \citenamefont {Pitaevskii},
  \citenamefont {Lifshitz},\ and\ \citenamefont {Sykes}}]{Landau1984}%
  \BibitemOpen
  \bibfield  {author} {\bibinfo {author} {\bibfnamefont {L.~D.}\ \bibnamefont
  {Landau}}, \bibinfo {author} {\bibfnamefont {J.}~\bibnamefont {Bell}},
  \bibinfo {author} {\bibfnamefont {M.}~\bibnamefont {Kearsley}}, \bibinfo
  {author} {\bibfnamefont {L.}~\bibnamefont {Pitaevskii}}, \bibinfo {author}
  {\bibfnamefont {E.}~\bibnamefont {Lifshitz}}, \ and\ \bibinfo {author}
  {\bibfnamefont {J.}~\bibnamefont {Sykes}},\ }\href@noop {} {\emph {\bibinfo
  {title} {Electrodynamics of continuous media}}},\ Vol.\ \bibinfo {volume} {8
  \S 26}\ (\bibinfo  {publisher} {Elsevier},\ \bibinfo {year}
  {1984})\BibitemShut {NoStop}%
\bibitem [{\citenamefont {Guyer}\ \emph {et~al.}(2009)\citenamefont {Guyer},
  \citenamefont {Wheeler},\ and\ \citenamefont {Warren}}]{FiPy2009}%
  \BibitemOpen
  \bibfield  {author} {\bibinfo {author} {\bibfnamefont {J.~E.}\ \bibnamefont
  {Guyer}}, \bibinfo {author} {\bibfnamefont {D.}~\bibnamefont {Wheeler}}, \
  and\ \bibinfo {author} {\bibfnamefont {J.~A.}\ \bibnamefont {Warren}},\
  }\href {\doibase 10.1109/MCSE.2009.52} {\bibfield  {journal} {\bibinfo
  {journal} {Computing in Science \& Engineering}\ }\textbf {\bibinfo {volume}
  {11}},\ \bibinfo {pages} {6} (\bibinfo {year} {2009})}\BibitemShut {NoStop}%
\bibitem [{\citenamefont {McIver}\ \emph {et~al.}(2011)\citenamefont {McIver},
  \citenamefont {Hsieh}, \citenamefont {Steinberg}, \citenamefont
  {Jarillo-Herrero},\ and\ \citenamefont {Gedik}}]{McIver2011}%
  \BibitemOpen
  \bibfield  {author} {\bibinfo {author} {\bibfnamefont {J.~W.}\ \bibnamefont
  {McIver}}, \bibinfo {author} {\bibfnamefont {D.}~\bibnamefont {Hsieh}},
  \bibinfo {author} {\bibfnamefont {H.}~\bibnamefont {Steinberg}}, \bibinfo
  {author} {\bibfnamefont {P.}~\bibnamefont {Jarillo-Herrero}}, \ and\ \bibinfo
  {author} {\bibfnamefont {N.}~\bibnamefont {Gedik}},\ }\href {\doibase
  10.1038/nnano.2011.214} {\bibfield  {journal} {\bibinfo  {journal} {Nat.
  Nanotechnol.}\ }\textbf {\bibinfo {volume} {7}},\ \bibinfo {pages} {96–100}
  (\bibinfo {year} {2011})}\BibitemShut {NoStop}%
\bibitem [{\citenamefont {Markvart}(2008)}]{Markvart2008}%
  \BibitemOpen
  \bibfield  {author} {\bibinfo {author} {\bibfnamefont {T.}~\bibnamefont
  {Markvart}},\ }\href {\doibase 10.1002/pssa.200880460} {\bibfield  {journal}
  {\bibinfo  {journal} {Phys. Status Solidi A}\ }\textbf {\bibinfo {volume}
  {205}},\ \bibinfo {pages} {2752–2756} (\bibinfo {year} {2008})}\BibitemShut
  {NoStop}%
\bibitem [{\citenamefont {Rau}(2007)}]{Rau2007}%
  \BibitemOpen
  \bibfield  {author} {\bibinfo {author} {\bibfnamefont {U.}~\bibnamefont
  {Rau}},\ }\href {\doibase 10.1103/physrevb.76.085303} {\bibfield  {journal}
  {\bibinfo  {journal} {Phys. Rev. B}\ }\textbf {\bibinfo {volume} {76}},\
  \bibinfo {pages} {085303} (\bibinfo {year} {2007})}\BibitemShut {NoStop}%
\end{thebibliography}
\end{document}